\documentclass{iopjournal}

\usepackage{booktabs}
\usepackage{float}

\hypersetup{colorlinks=true, urlcolor=blue, linkcolor=black, citecolor=blue}
\newcommand{\vollink}[2]{\href{#1}{{\color{blue}#2}}}
\let\OLDthebibliography\thebibliography
\renewcommand\thebibliography[1]{%
  \OLDthebibliography{#1}%
  \setlength{\itemsep}{0pt}
  \setlength{\parsep}{0pt}%
  \setlength{\parskip}{0pt}%
  \renewcommand{\baselinestretch}{0.93}
  \footnotesize
}

\begin{document}

\articletype{Paper}

\title{\texorpdfstring{\textcolor{blue}{Two-Dimensional Ferromagnetic Weyl Half-Semimetal}}{Two-Dimensional Ferromagnetic Weyl Half-Semimetal}}

\author{Yu-Yang Zhao$^{1,\#}$, Chen-Hao Lv$^{1,\#}$ Qing-Bo Liu$^{1,*}$, Wen-Neng Zhao$^1$, Xin-Yi Chen$^1$ and Lun Xiong$^{1,*}$}

\affil{$^1$Hubei Key Laboratory of Optical Information and Pattern Recognition, School of Optical Information and Energy Engineering, School of Mathematics and Physics, School of Chemical Engineering and Pharmacy, Wuhan Institute of Technology, Wuhan 430073, China}

\affil{$^*$Author to whom any correspondence should be addressed.}
\affil{$^\#$Authors are contributed equally to this work.}

\email{phyqboliu@wit.edu.cn; xionglun@wit.edu.cn}


\begin{abstract}
Two-dimensional (2D) ferromagnetic topological materials provide appealing platforms for realizing spin-polarized topological states, yet the coexistence of Weyl fermions and robust high-temperature ferromagnetism remains elusive. Here, based on first-principles calculations and Monte Carlo simulations, we predict monolayer BaMn$_4$O$_8$ to be a dynamically, thermally, and mechanically stable ferromagnetic Weyl half-semimetal. \textcolor{blue}{An anisotropic Monte Carlo simulation on a $20\times20\times1$ lattice yields a finite-size transition estimate of approximately 1446~K.} The spin-resolved electronic structure reveals a fully spin-polarized state, in which the spin-up channel hosts linear Weyl band crossings near the Fermi level, whereas the spin-down channel remains insulating. Corresponding Fermi-arc edge states connecting the projected Weyl points (WPs) are clearly observed along the (100) boundary, confirming the nontrivial topology. Importantly, both the Weyl points and their associated Fermi arcs remain robust in the presence of spin--orbit coupling (SOC). \textcolor{blue}{These results identify monolayer BaMn$_4$O$_8$ as a candidate for fully spin-polarized topological transport.}
\end{abstract}

\section{Introduction}

Weyl semimetals represent a fundamental class of topological semimetals characterized by symmetry-protected linear band crossings, known as Weyl points (WPs), between valence and conduction bands \cite{1,2,3,4}. In three dimensions, an isolated WP can be described by an effective two-band Weyl Hamiltonian and acts as a quantized source or sink of Berry curvature in momentum space \cite{5,6,7,8}. Accordingly, each WP carries a nonzero topological charge determined by the Berry-flux integral over a closed surface enclosing the node, while Weyl points with opposite chiralities are connected through characteristic Fermi-arc surface states as required by the bulk--boundary correspondence \cite{9}. Following the theoretical prediction and subsequent experimental realization of Weyl phases in noncentrosymmetric transition-metal monopnictides \cite{10,11}, considerable attention has been devoted to 2D topological semimetals. In contrast to their three-dimensional counterparts, isolated linear band crossings in 2D systems are commonly characterized by quantized Berry phases or winding numbers defined along closed momentum-space loops surrounding the nodes, and can be protected by crystalline symmetries \cite{12,13,14}. Their reduced dimensionality also provides enhanced tunability through strain, electrostatic gating, substrates, and external fields \cite{15,16,17,18}. When ferromagnetism is further introduced, the breaking of time-reversal symmetry lifts spin degeneracy and can generate strongly spin-polarized topological band crossings and associated boundary states \cite{19,20,21}. The combination of nontrivial topology, full spin polarization, and robust ferromagnetic order therefore makes 2D ferromagnetic Weyl semimetals particularly attractive for topological spintronics.

Motivated by this prospect, extensive efforts have been devoted to identifying 2D materials that integrate nontrivial band topology with intrinsic magnetism. Various Dirac- and Weyl-type states have been predicted in organometallic frameworks \cite{22}, manganese trihalides \cite{23}, and carbon allotropes \cite{24}, while Weyl points and associated boundary states have been reported in monolayer PtTe$_{1.75}$ \cite{25}. Related topological phenomena, including the quantum spin Hall effect in monolayer WTe$_2$ \cite{26,27}, topological superconductivity in ScSiX (X = Cl, Br) \cite{28}, and Fermi-arc states in layered electrides \cite{29}, further demonstrate the rich topological physics accessible in low-dimensional systems. In magnetic 2D materials, fully spin-polarized Weyl fermions and tunable quantum anomalous Hall phases have also been proposed \cite{30,31,32,33}. In parallel, substantial progress has been made toward enhancing the magnetic ordering temperature of 2D systems. Monolayer CrI$_3$ exhibits an intrinsic $T_{\mathrm C}$ of approximately 45~K \cite{43}, while atomically thin Fe$_3$GeTe$_2$ shows a $T_{\mathrm C}$ of 130--207~K depending on thickness \cite{44}, which can be further increased to room temperature by ionic gating \cite{45}. Ferromagnetic order above room temperature has been reported or predicted in VSe$_2$ \cite{46}, Fe$_3$ZnTe$_2$ and Fe$_3$ZnI$_2$ with $T_{\mathrm C}$ values of 390 and 560~K, respectively \cite{47}, and 2D FeB with a predicted $T_{\mathrm C}$ of approximately 750~K \cite{48}. 2D $\varepsilon$-Fe$_2$O$_3$ exhibits ferrimagnetic ordering up to approximately 800~K \cite{49}, whereas Fe$_{0.98}$In$_{0.02}$Te has been reported to exhibit a $T_{\mathrm C}$ as high as 860~K \cite{50}. Despite these advances, simultaneously realizing Weyl topology, full spin polarization, and robust ferromagnetic ordering above 1000~K in a single 2D material remains highly challenging.

To address this outstanding challenge, we employ first-principles calculations and Monte Carlo simulations to propose monolayer BaMn$_4$O$_8$ as a 2D \textcolor{blue}{ferromagnetic Weyl half-semimetal}. Its structural, dynamical, thermal, and mechanical stability is systematically verified through structural optimization, phonon calculations, ab initio molecular dynamics simulations, and elastic-constant analysis. Total-energy comparisons among different magnetic configurations establish the ferromagnetic ground state, \textcolor{blue}{while Monte Carlo simulations including magnetic anisotropy yield a transition estimate of approximately 1446~K on a $20\times20\times1$ lattice.} The spin-resolved electronic structure reveals a fully spin-polarized half-semimetallic state: the spin-up channel hosts linear Weyl crossings near the Fermi level, whereas the spin-down channel remains insulating. Consistent with the bulk--boundary correspondence, clear Fermi-arc edge states connecting the projected Weyl points emerge along the (100) boundary, confirming the nontrivial band topology. Importantly, both the Weyl points and the associated Fermi arcs remain robust in the presence of SOC. \textcolor{blue}{Monolayer BaMn$_4$O$_8$ thus combines a ferromagnetic ground state with fully spin-polarized Weyl crossings and topological edge states.}

\section{Computational methods}
\begin{figure}  
\includegraphics[width=0.95\textwidth]{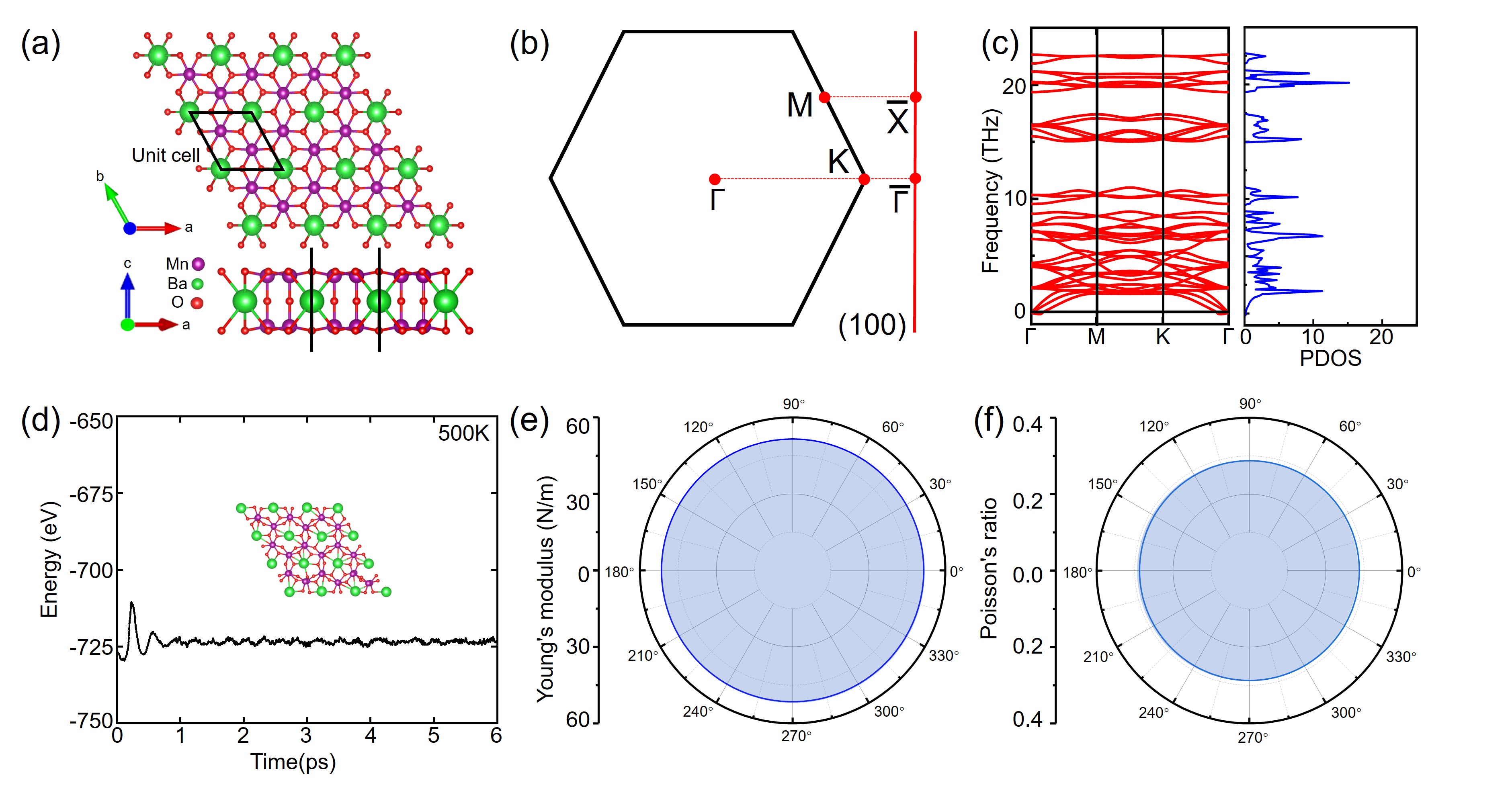}
\caption{(a) Top and side views of the monolayer BaMn$_4$O$_8$, with 1 Ba atom, 4 Mn atoms, and 8 O atoms in a unit cell.  (b) Projection of the 2D Brillouin zone onto the one-dimensional Brillouin zone of the (100) edge. (c) Phonon dispersion and density of states of 2D BaMn$_4$O$_8$. (d) AIMD stability of the 2D BaMn$_4$O$_8$ at 500~K using a $(3\times3\times1)$ supercell. (e) Young's modulus and (f) Poisson's ratio.}
\label{fig:1}
\end{figure}
We perform spin-polarized first-principles calculations within the framework of density functional theory (DFT) \cite{51,52}, as implemented in the Vienna \textit{Ab initio} Simulation Package (\textsc{VASP}) \cite{53,54}. The projector-augmented-wave (PAW) method is employed to describe the electron--ion interactions, while the exchange--correlation energy is treated using the Perdew--Burke--Ernzerhof (PBE) functional within the generalized gradient approximation (GGA) \cite{55}. The on-site Coulomb interaction of the Mn $3d$ electrons is described using the Liechtenstein formulation of the PBE+$U$ method \cite{56}, with the effective interaction parameters $U=3.9$~eV and $J=0$~eV adopted for the Mn $3d$ states following previous DFT+$U$ work on Mn-based oxides \cite{57}. \textcolor{blue}{The robustness of the predicted Weyl half-semimetallic state against the choice of $U$ has been verified by additional spin--orbit-coupled band-structure calculations with $U=3.3$ and $4.5$~eV.} A plane-wave energy cutoff of 520~eV and a $\Gamma$-centered $6\times6\times1$ $k$-point mesh are adopted. A vacuum layer of 20~\AA{} is introduced along the out-of-plane direction to eliminate interactions between periodic images. The structure is fully relaxed until the residual Hellmann--Feynman forces on each atom \textcolor{blue}{are below} $0.01~\mathrm{eV/\AA}$, and the total-energy convergence criterion \textcolor{blue}{is set to} $10^{-6}$~eV. Maximally localized Wannier functions are constructed using \textsc{Wannier90} \cite{58}, and the corresponding tight-binding Hamiltonian is employed in \textsc{WannierTools} \cite{59} to calculate the edge spectral functions and Fermi-arc states. The irreducible representations of the electronic bands are calculated using IRVSP \cite{58_1} program on the electronic Hamiltonian of the TB model and the low-energy $\mathbf{k}\cdot\mathbf{p}$ Hamiltonians are constructed using VASP2KP \cite{58_2}. Phonon dispersions \textcolor{blue}{are calculated} using density functional perturbation theory as implemented in \textsc{VASP} and postprocessed with \textsc{Phonopy} \cite{60}. The magnetic exchange parameters \textcolor{blue}{are extracted} by mapping the first-principles total energies of different magnetic configurations onto a classical Heisenberg model.

\textcolor{blue}{Classical Monte Carlo simulations are performed with the VAMPIRE code \cite{61}, using the Metropolis algorithm with adaptive spin moves \cite{62}. The simulation uses a $20\times20\times1$ lattice containing 1600 Mn spins and runs for 100,000 Monte Carlo steps. Magnetic anisotropy is included in the spin Hamiltonian.} \textcolor{blue}{Ab initio molecular dynamics (AIMD) simulations are performed for 5 ps (5000 steps) with a time step of 1 fs in a 3$\times$3$\times$1 supercell at 1400~K using the NVT ensemble with the Nos\'e--Hoover thermostat to control temperature.}

\section{RESULTS AND DISCUSSION}

\begin{figure}
\includegraphics[width=0.95\textwidth]{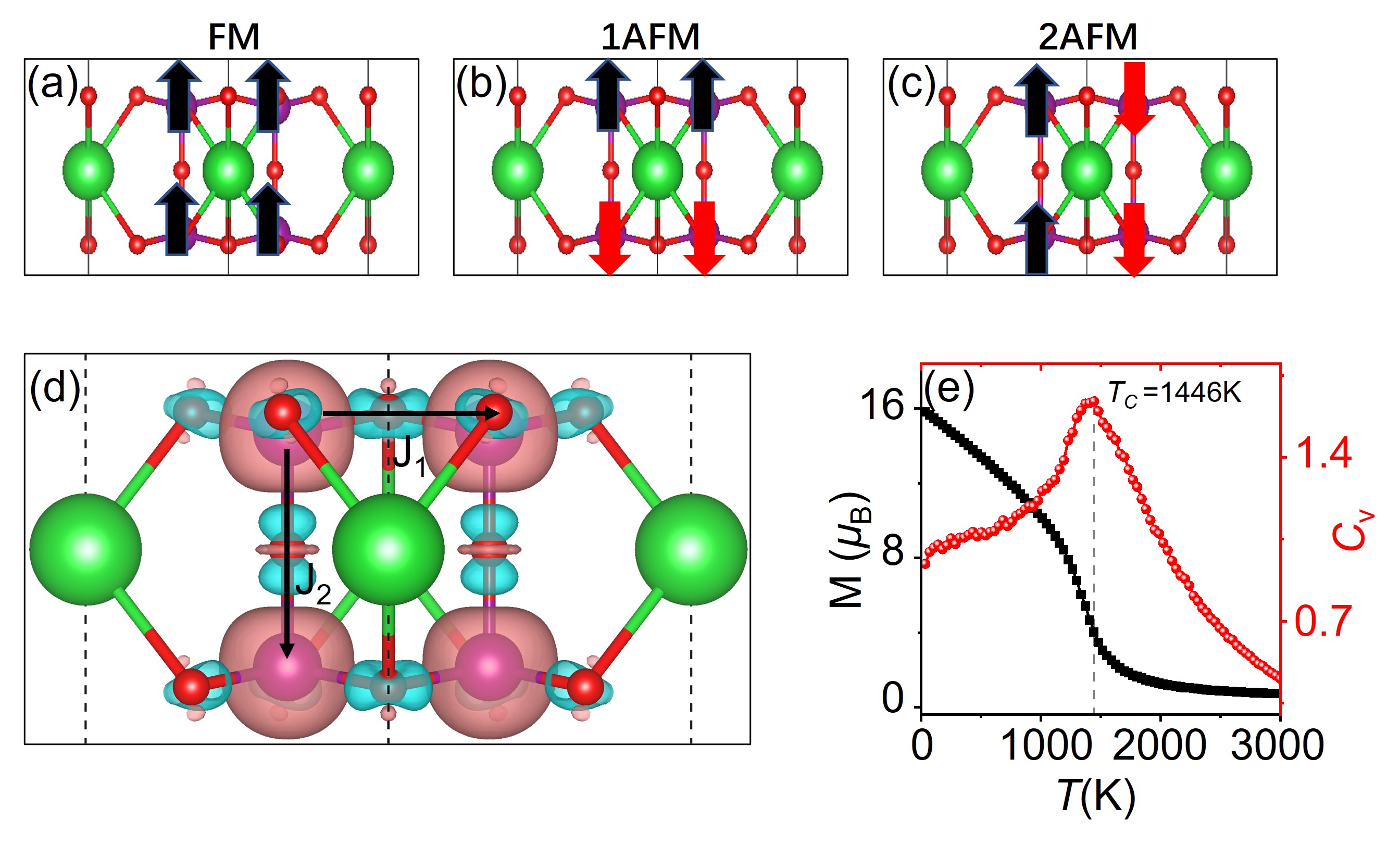}
\caption{(a--c) Magnetic configurations of BaMn$_4$O$_8$ in the FM, 1AFM, and 2AFM states, where the black (red) arrows represent the spin-up (-down). (d) Two magnetic exchange paths $J_1$ and $J_2$, where \textcolor{blue}{$J_1$ and $J_2$ denote the nearest- and next-nearest-neighbor Mn-Mn exchange interactions, respectively.} \textcolor{blue}{(e) Temperature dependence of the out-of-plane Mn magnetization and specific heat for the $20\times20\times1$ lattice. The dashed vertical line marks the finite-size transition estimate of approximately 1446~K.}}
\label{fig:2}
\end{figure}

\begin{figure}
\centering
\includegraphics[width=0.95\textwidth]{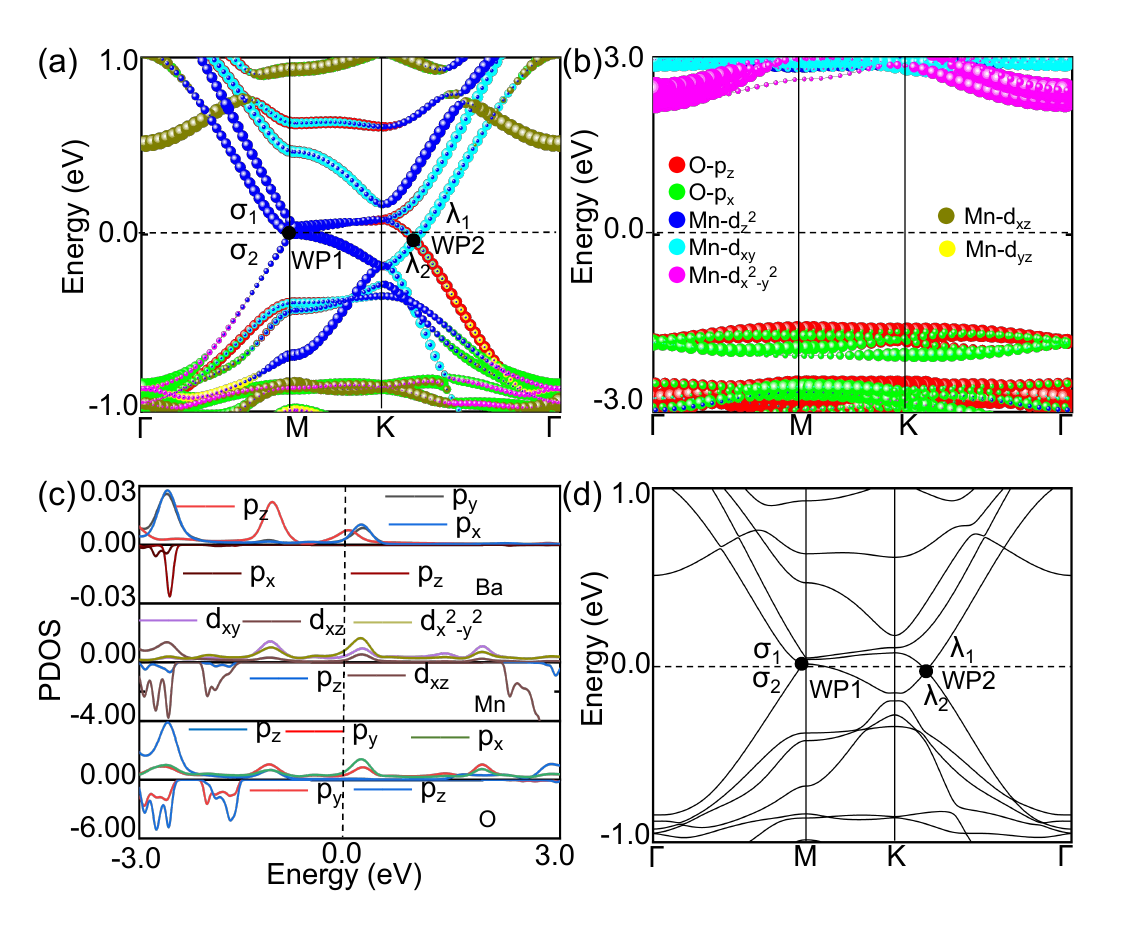}
\caption{Spin-resolved electronic structures of ferromagnetic monolayer BaMn$_4$O$_8$. (a) Spin-up band structure, where WP1 and WP2 mark the Weyl points with crossing bands of irreps
$\sigma_1 \oplus \sigma_2$ and $\lambda_1 \oplus \lambda_2$. (b) Spin-down band structure. (c) Orbital-projected density of states. (d) Band structure with SOC.}
\label{fig:3*}
\end{figure}

\subsection{Crystal structures and mechanical properties}

The predicted BaMn$_4$O$_8$ monolayer crystallizes in a hexagonal lattice with type-III magnetic space group $P\bar{3}1m'$ (BNS No. 162.77; OG No. 162.5.1307), as shown in Fig.~\ref{fig:1}(a). Each Ba atom is coordinated by six O atoms, while each Mn atom is surrounded by six O atoms, forming a MnO$_6$ coordination environment. The average electrostatic potential along the z-axis and the electron localization function (ELF) in Fig.~S1 of Supplemental Material (SM) show pronounced electron accumulation around the O atoms and depletion around the Ba and Mn atoms, indicating predominantly ionic Ba--O and Mn--O bonding. This bonding character is further corroborated by Bader charge analysis \cite{63}, which reveals substantial charge transfer from Ba and Mn to O. To facilitate the analysis of the edge topology, Fig.~\ref{fig:1}(b) shows the projection of the 2D Brillouin zone onto the one-dimensional Brillouin zone associated with the (100) edge.\textcolor{blue}{Under this projection, the $\Gamma$ point and the $M$ point at the top of the Brillouin-zone boundary are mapped onto $\bar{\Gamma}$ and $\bar{X}$, respectively.}This projected Brillouin zone is subsequently employed to analyze the edge spectral functions and Fermi-arc states along the (100) direction.

The stability of 2D BaMn$_4$O$_8$ was systematically examined from dynamical, thermal, and mechanical perspectives. As shown in Fig.~\ref{fig:1}(c), the phonon spectrum exhibits no imaginary modes throughout the Brillouin zone, confirming its dynamical stability. Furthermore, \textit{ab initio} molecular dynamics simulations at 500~K show that the monolayer retains its structural integrity, with the total energy fluctuating around a stable average value throughout the 6~$ps$ simulation [Fig.~\ref{fig:1}(d)], supporting its thermal stability. Additional AIMD simulations at 100, 200, 300, and 400 K, shown in Fig. S2 of the SM, exhibit similarly small energy fluctuations without noticeable structural reconstruction. \textcolor{blue}{Moreover, ab initio molecular dynamics simulations at 1400~K confirm that the structural framework remains intact (Fig.~S3 of the SM), and the thermodynamic stability against competing Ba--Mn--O phases is assessed via the ternary convex hull (Fig.~S4 of the SM), where BaMn$_4$O$_8$ lies only 28.5~meV per atom above the hull, indicating reasonable thermodynamic stability.} The calculated elastic constants, $C_{11}=56.15$~N/m and $C_{12}=16.144$~N/m, satisfy the Born mechanical stability criteria for a hexagonal 2D lattice, namely $C_{11}>0$ and $C_{11}>|C_{12}|$. The corresponding in-plane Young's modulus and Poisson's ratio are 51.5~N/m and 0.29, respectively, indicating that the monolayer is mechanically stable while retaining appreciable flexibility [Figs.~\ref{fig:1}(e) and \ref{fig:1}(f)].

\begin{table}[t]
\centering
\caption{Total energies of monolayer BaMn$_4$O$_8$ in the ferromagnetic (FM) and two antiferromagnetic (1AFM and 2AFM) configurations, calculated using the PBE+$U$ method.}
\label{tab:magnetic_energies}
\begin{tabular}{lccc}
\toprule
Configuration & FM & 1AFM & 2AFM \\
\midrule
$E_{\mathrm{tot}}$ (eV) & $-90.3301$ & $-89.1859$ & $-88.4163$ \\
\bottomrule
\end{tabular}
\end{table}

\subsection{\texorpdfstring{\textcolor{blue}{Magnetic order and finite-size transition estimate}}{Magnetic order and finite-size transition estimate}}

To determine the magnetic ground state of monolayer BaMn$_4$O$_8$, we considered one ferromagnetic configuration (FM) and two antiferromagnetic configurations (1AFM and 2AFM), as illustrated in Figs.~2(a)--2(c). The total-energy comparison shows that the FM configuration is energetically more favorable than both AFM states (Table~\ref{tab:magnetic_energies}), establishing the ferromagnetic ground state. The four Mn atoms in the unit cell collectively carry a magnetic moment of approximately $16~\mu_{\mathrm B}$, corresponding to about $4~\mu_{\mathrm B}$ per Mn atom.

To clarify the microscopic origin of the ferromagnetic ordering, we further examined the magnetic exchange interactions between neighboring Mn atoms. \textcolor{blue}{We describe the Mn--Mn exchange using a classical Heisenberg model, with positive exchange constants denoting ferromagnetic coupling.} As shown in Fig.~2(d), two inequivalent Mn--O--Mn superexchange pathways are characterized by $J_1=0.03987$~eV and $J_2=0.07152$~eV. \textcolor{blue}{The robustness of the exchange hierarchy $J_2>J_1$ has been verified by fitting the exchange constants from a $2\times2\times1$ supercell with four magnetic configurations (FM, 1AFM, stripe, and zigzag), yielding $J_1=0.0463$~eV and $J_2=0.0802$~eV (Table~S1), in agreement with the primitive-cell results in sign and hierarchy. The third-nearest-neighbour diagonal Mn--Mn path (4.93~\AA{}) lacks a common bridging oxygen and contributes an exchange below 1~meV, confirming that no significant exchange channel is omitted.}\textcolor{blue}{The dominance of $J_2$ has a clear geometric origin: the interlayer path proceeds through a nearly linear Mn--O--Mn bridge (bond angle 180.0$^\circ$), which maximizes the coherent $t_{2g}$--$p$--$t_{2g}$ hopping, whereas the in-plane path has an intermediate angle of 121.7$^\circ$.} Within the adopted Heisenberg Hamiltonian convention, the positive values of both exchange parameters indicate ferromagnetic coupling. Moreover, $J_2>J_1>0$ indicates that the second exchange pathway provides the stronger ferromagnetic interaction and therefore plays a dominant role in stabilizing the FM ground state. \textcolor{blue}{The magnetic anisotropy energy, defined as $E_z-E_x$, is $-30~\mu\mathrm{eV}$ per unit cell, indicating an out-of-plane easy axis.}

{\color{blue}
A minimal spin Hamiltonian for the two exchange paths and the out-of-plane anisotropy is
\begin{equation}
\label{eq:spin-hamiltonian}
\mathcal{H} =
-J_1 \sum_{\langle i,j\rangle_1} \mathbf{s}_i\cdot\mathbf{s}_j
-J_2 \sum_{\langle i,j\rangle_2} \mathbf{s}_i\cdot\mathbf{s}_j
-K \sum_i (s_i^z)^2 .
\end{equation}
Here $\mathbf{s}_i$ is a unit vector along the local Mn spin direction, $s_i^z$ is its out-of-plane component, and each Mn--Mn bond in a given exchange path is counted once. In the uniform second-order single-ion approximation, the calculated anisotropy energy is shared equally among the four Mn sites in the unit cell. This gives $E_z-E_x=-4K$ and hence $K=+7.5~\mu\mathrm{eV}$ per Mn. The positive $K$ favors out-of-plane magnetization. The exchange favors parallel alignment, while the easy-axis anisotropy suppresses long-wavelength transverse spin fluctuations \cite{torelli2019critical}.
\par}

\textcolor{blue}{For the $20\times20\times1$ lattice, the specific heat $C_V$ peaks near 1446~K, where the out-of-plane magnetization $M_z$ decreases to nearly zero [Fig.~\ref{fig:2}(e)]. We therefore report approximately 1446~K as the finite-size transition estimate for this lattice. Determining the Curie temperature in the thermodynamic limit requires simulations at several lateral sizes. One can test convergence by tracking the crossings of the Binder cumulant $U_4=1-\langle M_z^4\rangle/(3\langle M_z^2\rangle^2)$ for different sizes and comparing their evolution with that of the specific-heat and susceptibility peaks \cite{binder1981scaling,binder1976heisenberg}.}

\subsection{Nontrivial topological properties}

We next examine the electronic and topological properties of the ferromagnetic ground state. The spin-resolved band structures and orbital-projected density of states of monolayer BaMn$_4$O$_8$ are shown in Fig. 3. The spin-up channel exhibits semimetallic behavior, with two linear band crossings located near the Fermi level $E_{\mathrm F}$ [Fig. 3(a)], whereas the spin-down channel is insulating with a band gap of 3.869~eV [Fig. 3(b)]. The low-energy states around the Fermi level mainly originate from the hybridization of the Mn-$d_{x^{2}-y^{2}}$, Mn-$d_{xy}$, O-$p_x$, and O-$p_y$ orbitals, as evidenced by the projected density of states in Fig. 3(c). The coexistence of a semimetallic majority-spin channel and an insulating minority-spin channel therefore gives rise to a fully spin-polarized half-semimetallic state. By comparison, the antiferromagnetic configuration does not exhibit such Weyl half-semimetallic behavior, as shown in Fig.~S5 of SM.
\begin{figure}
\includegraphics[width=0.95\textwidth]{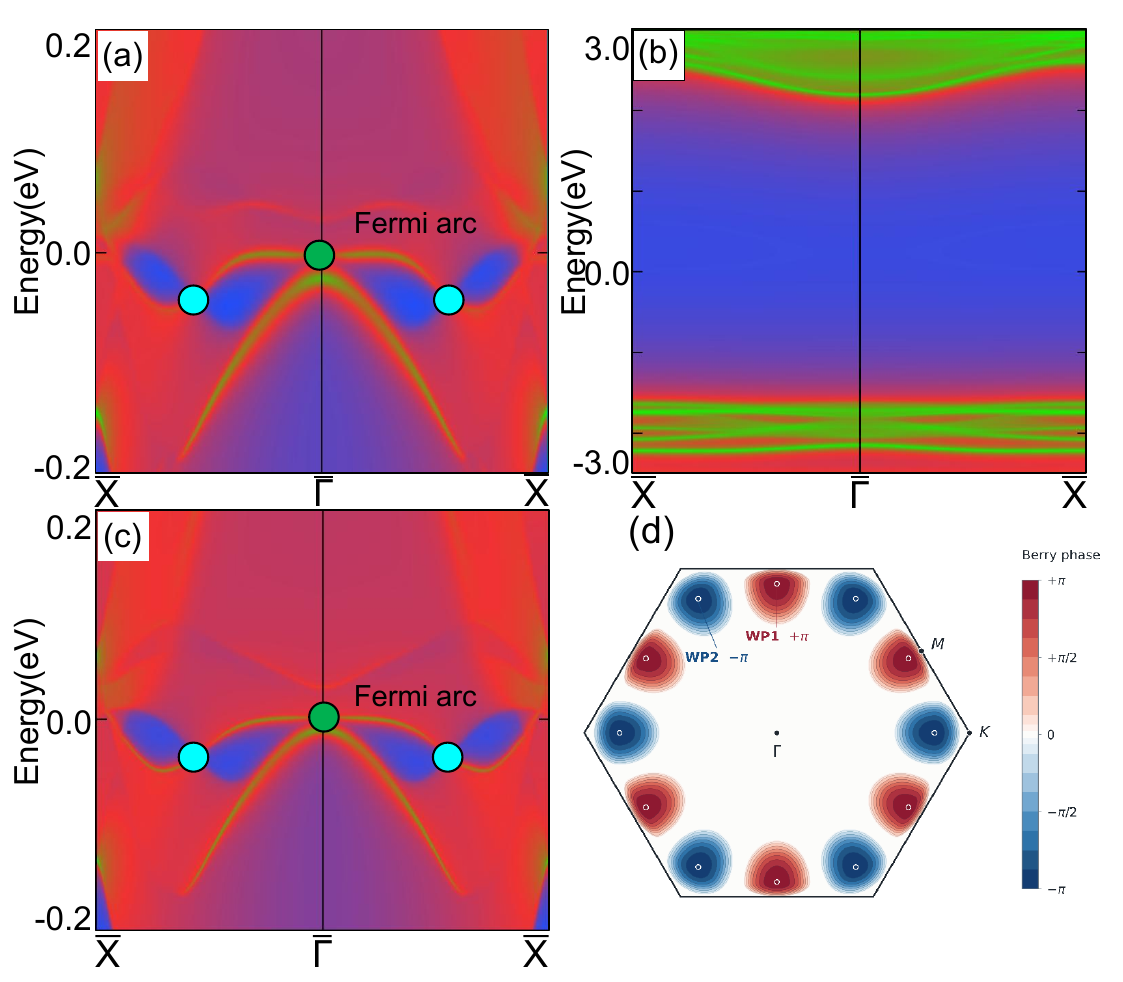}
\caption{(a--b) Edge spectral functions of the (100) boundary for the spin-up and spin-down channels, respectively. (c) Edge spectral function of the (100) boundary with SOC. {\color{blue}(d) Weyl-point distribution and Berry phases in the 2d BZ. Open circles mark the Weyl points. The Berry phases are +$\pi$ for WP1 (red) and -$\pi$ for WP2 (blue), as indicated by the color scale.}}
\end{figure}
A closer examination of the spin-up bands identifies two Weyl points, denoted WP1 and WP2, located at
$\mathbf{k}_{WP1}=(0.488597,0,0)$ along the $\Gamma$--$M$ direction and
$\mathbf{k}_{WP2}=(0.273843,0.273843,0)$ along the $\Gamma$--$K$ direction in reciprocal-lattice coordinates, respectively. Symmetry analysis shows that the two crossing bands at WP1 transform according to the distinct irreducible representations $\sigma_1 \oplus \sigma_2$, while those at WP2 belong to $\lambda_1 \oplus \lambda_2$. Because bands belonging to different irreducible representations cannot hybridize along the corresponding high-symmetry lines, the two linear crossings are symmetry protected.

To further characterize the low-energy dispersions around these Weyl points, we constructed two-band $\mathbf{k}\cdot\mathbf{p}$ Hamiltonians. Defining $\mathbf{q}=\mathbf{k}-\mathbf{k}_{WP1}$, the effective Hamiltonian around WP1 can be written as
\begin{equation}
H_{WP1}^{\mathbf{k}\cdot\mathbf{p}}(\mathbf{q})=
\left(
\begin{array}{cc}
a_1+(b_1+b_4)u_1 & (-b_2+ib_3)v_1 \\
(-b_2-ib_3)v_1 & a_1+(b_1-b_4)u_1
\end{array}
\right),
\end{equation}
where
$u_1=\sqrt{3}q_x+q_y$ and
$v_1=\frac{\sqrt{3}}{3}q_x-q_y$.

Similarly, for WP2, with $\mathbf{q}=\mathbf{k}-\mathbf{k}_{WP2}$, the effective Hamiltonian reads
\begin{equation}
H_{WP2}^{\mathbf{k}\cdot\mathbf{p}}(\mathbf{q})=
\left(
\begin{array}{cc}
c_1+(d_1+d_4)u_2 & (-d_2+id_3)v_2 \\
(-d_2-id_3)v_2 & c_1+(d_1-d_4)u_2
\end{array}
\right),
\end{equation}
where
$u_2=\frac{\sqrt{3}}{3}q_x+q_y$ and
$v_2=\sqrt{3}q_x-q_y$.
The $a_1$, $b_n$ (n=1, 2, 3, 4), $c_1$, and $d_n$ (n=1, 2, 3, 4) are constants. The fitted parameters are summarized in Table~\ref{tab:kp_params}, and the corresponding $\mathbf{k}\cdot\mathbf{p}$ dispersions are compared with the DFT bands in Fig.~S6 of SM. The good agreement between the two descriptions confirms the linear and anisotropic dispersions around both WP1 and WP2.

\begin{table}[h!]
\centering
\caption{\textcolor{blue}{Fitted $\mathbf{k}\cdot\mathbf{p}$ parameters for the
Weyl points WP1 and WP2 in monolayer BaMn$_4$O$_8$.}}
\label{tab:kp_params}
\begin{tabular}{cccc}
\toprule
\multicolumn{2}{c}{WP1 ($\Gamma$--$M$)} & \multicolumn{2}{c}{WP2 ($\Gamma$--$K$)} \\
\cmidrule(lr){1-2} \cmidrule(lr){3-4}
Parameter & \textcolor{blue}{Value} & Parameter & \textcolor{blue}{Value} \\
\midrule
$a_1$ \textcolor{blue}{(eV)} & $-3.3941$ & $c_1$ \textcolor{blue}{(eV)} & $-3.4327$ \\
$b_1$ \textcolor{blue}{(eV·Å)} & $\phantom{-}0.4074$ & $d_1$ \textcolor{blue}{(eV·Å)} & $-0.0431$ \\
$b_2$ \textcolor{blue}{(eV·Å)} & $\phantom{-}0.1541$ & $d_2$ \textcolor{blue}{(eV·Å)} & $-0.0365$ \\
$b_3$ \textcolor{blue}{(eV·Å)} & $\phantom{-}0.5707$ & $d_3$ \textcolor{blue}{(eV·Å)} & $-1.3498$ \\
$b_4$ \textcolor{blue}{(eV·Å)} & $\phantom{-}0.0117$ & $d_4$ \textcolor{blue}{(eV·Å)} & $-0.0197$ \\
\bottomrule
\end{tabular}
\end{table}

We further examined the robustness of these Weyl points against SOC. As shown in Fig. 3(d), the low-energy band crossings remain nearly unchanged after SOC is included, with only minor modifications to the surrounding band dispersions. \textcolor{blue}{The persistence of the crossings under SOC is consistent with the magnetic-symmetry analysis: the ferromagnetic state is described by the magnetic space group $P\bar{3}1m'$ (No.~162.77, BNS), and the little co-group of the $\Gamma$--$M$ line is $m'$ with one-dimensional co-representations, so the crossings are protected by their topological charge rather than by a symmetry-enforced degeneracy. The corresponding $\mathbf{k}\cdot\mathbf{p}$ Hamiltonians are given in the SM. Figure~S7 compares the fitted and DFT dispersions, and Table~S1 lists the fitted parameters. In addition, spin--orbit-coupled band-structure calculations with $U=3.3$ and $4.5$~eV confirm that the Weyl half-semimetallic state is robust against the choice of $U$ (Fig.~S8 of the SM).}

Finally, to reveal the boundary manifestation of the nontrivial bulk band topology, we calculated the edge spectral functions for the (100) boundary. As shown in Figs.~4(a)--4(c), pronounced Fermi-arc edge states emerge near the Fermi level and connect the projected bulk Weyl points in the one-dimensional edge Brillouin zone. Importantly, these edge states persist after SOC is included, further demonstrating the robustness of the nontrivial topology. \textcolor{blue}{We have further verified that the edge states computed for the (010) boundary are qualitatively identical to those of the (100) boundary (Fig.~S9 of the SM), confirming that the edge states originate from the bulk topology rather than from edge-specific accidental states.} As shown in Fig.~4(d), the Weyl points form symmetry-related sets in the first Brillouin zone, with their momentum coordinates summarized in Table~S2 of SM. Their distribution is constrained by the threefold rotational, inversion, and relevant magnetic symmetries of the ferromagnetic state. {\color{blue}The calculated Berry phases are +$\pi$ around WP1 and -$\pi$ around WP2 [Fig. 4(d)]} Taken together, the fully spin-polarized Weyl crossings, their symmetry-protected low-energy dispersions, and the associated Fermi-arc edge states establish monolayer BaMn$_4$O$_8$ as a 2D ferromagnetic Weyl half-semimetal.

\section{Conclusion}
In summary, we systematically investigated the structural stability, magnetic properties, and nontrivial band topology of monolayer BaMn$_4$O$_8$ using first-principles calculations and Monte Carlo simulations. Monolayer BaMn$_4$O$_8$ is identified as a ferromagnetic Weyl half-semimetal, featuring fully spin-polarized Weyl crossings near the Fermi level and corresponding Fermi-arc edge states along the (100) boundary. Both the Weyl points and the associated edge states remain robust in the presence of SOC, further supporting the robustness of the nontrivial topological state.  \textcolor{blue}{Monte Carlo simulations including magnetic anisotropy yield a finite-size transition estimate of approximately 1446~K for the $20\times20\times1$ lattice. The thermodynamic-limit Curie temperature remains to be determined by finite-size scaling. Monolayer BaMn$_4$O$_8$ is a candidate for spin-polarized topological transport.}


\funding{This work was supported by the National \textcolor{blue}{Natural} Science Foundation of China (Grant No. 12404085), \textcolor{blue}{by} the 2025 Hubei Provincial Key Laboratory Directed Commission Project (Basic Research Platform Program, Grant No. 2025CSA090), \textcolor{blue}{by} the Youth Talent Project of the Scientific Research Program of the Education Department of Hubei Province (\textcolor{blue}{Grant} No. Q20251505), \textcolor{blue}{by} the Graduate Innovative Fund of Wuhan Institute of Technology (\textcolor{blue}{Grant} No. CX2025276), \textcolor{blue}{and by} the Shiyan Key Laboratory of Quantum Information and Precision Optics (Hubei University of Automotive Technology, (\textcolor{blue}{Grant No.} SYZDK12024B03{)}).}

\data{The data that support the findings of this study are available from the corresponding author upon reasonable request.}

\suppdata{See the supplementary data for the stability analysis, AIMD simulations, antiferromagnetic band structures, $\mathbf{k}\cdot\mathbf{p}$ fittings of the Weyl points, and the complete list of Weyl-point positions of BaMn$_4$O$_8$.}

\end{document}